\documentclass[prl,amsmath,amssymb,twocolumn,superscriptaddress]{revtex4-2}

\usepackage{graphicx}
\usepackage{bm}
\usepackage[caption=false]{subfig}
\usepackage{color}
\usepackage{hyperref}  
\usepackage{soul}

\begin{document}

\title{Chiral Phonons in Moir\'e Superlattices} 
\author{Nishchay Suri}
\affiliation{Department of Physics, Carnegie Mellon University, Pittsburgh, Pennsylvania 15213, USA}

\author{Chong Wang}
\affiliation{Department of Materials Science and Engineering, University of Washington, Seattle, Washington 98195, USA}

\author{Yinhan Zhang}
\affiliation{Department of Physics, Carnegie Mellon University, Pittsburgh, Pennsylvania 15213, USA}

\author{Di Xiao}
\affiliation{Department of Materials Science and Engineering, University of Washington, Seattle, Washington 98195, USA}
\affiliation{Department of Physics, University of Washington, Seattle, Washington 98195, USA}
\email{dixiao@uw.edu}
\date{\today}

\begin{abstract}
We discover chiral phonons at the lowest-energy bands in moir\'e superlattices. The moir\'e chiral phonons we uncover are the collective excitations of the stacking domains. Their origin is uniquely attributed to the stacking configurations whose interlayer binding energy breaks the $C_{2z}$ symmetry on the moir\'e length scale.
Within elastic theory, we use a general symmetry analysis to provide a complete classification of van der Waals heterostructures in respect to hosting moir\'e chiral phonons and show the calculation for twisted MoS$_2$ as an example.
We present a low-energy effective model to qualitatively understand the moir\'e chiral phonons and show that it captures the essential physics remarkably well.
Our result potentially opens up new possibilities in phononic twistronics as the moir\'e chiral phonons have high tunability, moir\'e scale wavelengths, excitation energies in only a few meV, and can possibly be mechanically excited.
\end{abstract}

\maketitle

When two layers of van der Waals crystals with a small lattice mismatch or rotation are stacked together, a long-wavelength moir\'e pattern will emerge.  The moir\'e pattern acts like a superlattice potential on charge carriers, which can give rise to a variety of spectacular quantum phenomena~\cite{andrei2020,balents2020,kennes2021}.  The moir\'e superlattice can also significantly modify the phonon properties, leading to the formation of moir\'e minibands and strong renormalization of the speed of sound~\cite{koshino2019,ochoa2019,maity2020,lamparski2020,gadelha2021,quan2021}.  The modification of the low-energy acoustic phonon modes has important implications in thermal transport and electron-phonon interactions in van der Waals heterostructures~\cite{wu2018,lian2019,wu2019,koshino2020}.

On the other hand, it has been recently realized that phonons can acquire finite angular momentum and Berry curvature if inversion symmetry is broken~\cite{zhang2015}.  These phonons are called chiral phonons.  The angular momentum of chiral phonons decides selection rules in the electronic intervalley scattering~\cite{zhang2015,zhu2018,chen2019,he2020,liu2020,delhomme2020}, and their Berry curvature can lead to topological phonon transport~\cite{li2021}.  So far, studies of chiral phonons has been focused on crystals with broken sublattice symmetry\cite{zhang2015}. The resulting chiral phonon modes are high energy excitations with wavelengths on the order of atomic lattice constant and therefore, they can only be optically excited. 

In this Letter, we show that a new type of chiral phonon can emerge in moir\'e superlattices in van der Waals heterostructures. 
The moir\'e chiral phonons we discovered represent the collective motion of the stacking domains. The moir\'e superlattice causes acoustic branches of the monolayer phonon spectrum to form minibands. We show that these minibands carry chiral phonons that have moir\'e-scale wavelengths and can possibly be mechanically excited. They originate from stacking configurations whose interlayer binding energy breaks $C_{2z}$ symmetry on the scale of the moir\'e superlattice. Within the framework of elastic theory, we first present a general symmetry analysis to provide a complete classification of van der Waals heterostructures according to the existence or absence of chiral moir\'e phonons.  We then demonstrate the existence of these low-energy chiral phonons using twisted bilayer MoS$_2$ with the twist angle close to 180$^\circ$ as an example.  The chiral nature of these phonons is shown by explicit calculation of the phonon angular momentum and Berry curvature, as well as direct visualization of the real-space motion of the phonon modes.  To intuitively understand the moir\'e chiral phonons, we develop a low energy effective model where we show that the interlayer binding potential introduces an antisymmetric term in the effective dynamical matrix which leads to the chirality.  Finally, we show the tunability of the moir\'e chiral phonons by twist angle. The high tunability, excitation energies in only a few meV, superlattice scale wavelengths and possible mechanical excitation of the moir\'e chiral phonons potentially opens new possibilities in phononic twistronic devices.

Chiral phonons are characterized by their angular momentum $\bm L(\bm k)$.  For two-dimensional systems, it is sufficient to consider the $z$ component of $\bm L(\bm k)$.  If time-reversal symmetry is present, $L^z(\bm k)$ has the property $L^z(\bm k) = -L^z(-\bm k)$.  If inversion symmetry is also respected, $L^z(\bm k) =  L^z(-\bm{k})$, and $L^z(\bm k)$ vanishes.  Therefore, it is necessary to break inversion symmetry for the emergence of chiral phonons in non-magnetic systems.  The inversion symmetry can be further decomposed into the product of a two-fold rotation $C_{2z}$ around the $z$ axis and a mirror reflection $M_z$ with respect to the $xy$-plane.  It is straightforward to show that $C_{2z}$ also forbids $L^z(\bm k)$ but $M_z$ does not impose any symmetry constraint.

We now apply the above symmetry consideration to moir\'e phonons.  Since we are interested in low-energy phonons on the length scale of the moir\'e superlattice, we shall adopt the continuum approach based on standard elastic theory~\cite{carr2018,koshino2019,ochoa2019}.  Furthermore, only in-plane vibrations will be considered because out-of-plane vibrations do not contribute to $L_z(\bm k)$, and the corresponding flexural phonon modes were shown to remain nearly unchanged by the moir\'e superlattice~\cite{koshino2019,quan2021}.

The total energy of the system can be separated into an intralayer part and an interlayer part.  Introducing the in-plane deformation field $\bm u^{(\ell)}$ where $\ell \in \{1, 2\}$ is the layer index, the intralayer elastic energy can be expressed as~\cite{carr2018,koshino2019,ochoa2019}
\begin{equation}
\begin{split}
U_E = \sum_{\ell = 1}^2\int d^2r \,
\Bigl[ \frac{\lambda^{(\ell)}}{2}\bigl(\partial_\alpha u_\alpha^{(\ell)}\bigr)^{2} 
+ \frac{\mu^{(\ell)}}{4}\bigl(\partial_\alpha u_\beta^{(\ell)}
+ \partial_\beta u_\alpha^{(\ell)}\bigr)^{2} \Bigr] \;,
\end{split}
\end{equation}
where $\mu^{(\ell)}$ and $\lambda^{(\ell)}$ are the Lam\'e coefficients of layer $\ell$.   One can see that $U_E$ does not break $C_{2z}$~\footnote{Materials like monolayer hexagonal boron nitride(h-BN) indeed break inversion symmetry on atomic scale.  However, the theory of elasticity is a macroscopic continuum theory concerned with distances much larger than the atomic scale.  It provides a useful and correct description of the low energy and long-wavelength acoustic vibrations where the variation in deformation is smooth and small as compared to the distances. It is therefore oblivious to inversion breaking on the atomic scale.}.  It can break $M_z$ if the two layers are different, but the existence of $C_{2z}$ symmetry still forbids the appearance of chiral phonons.

\begin{figure}
\includegraphics[width=\columnwidth]{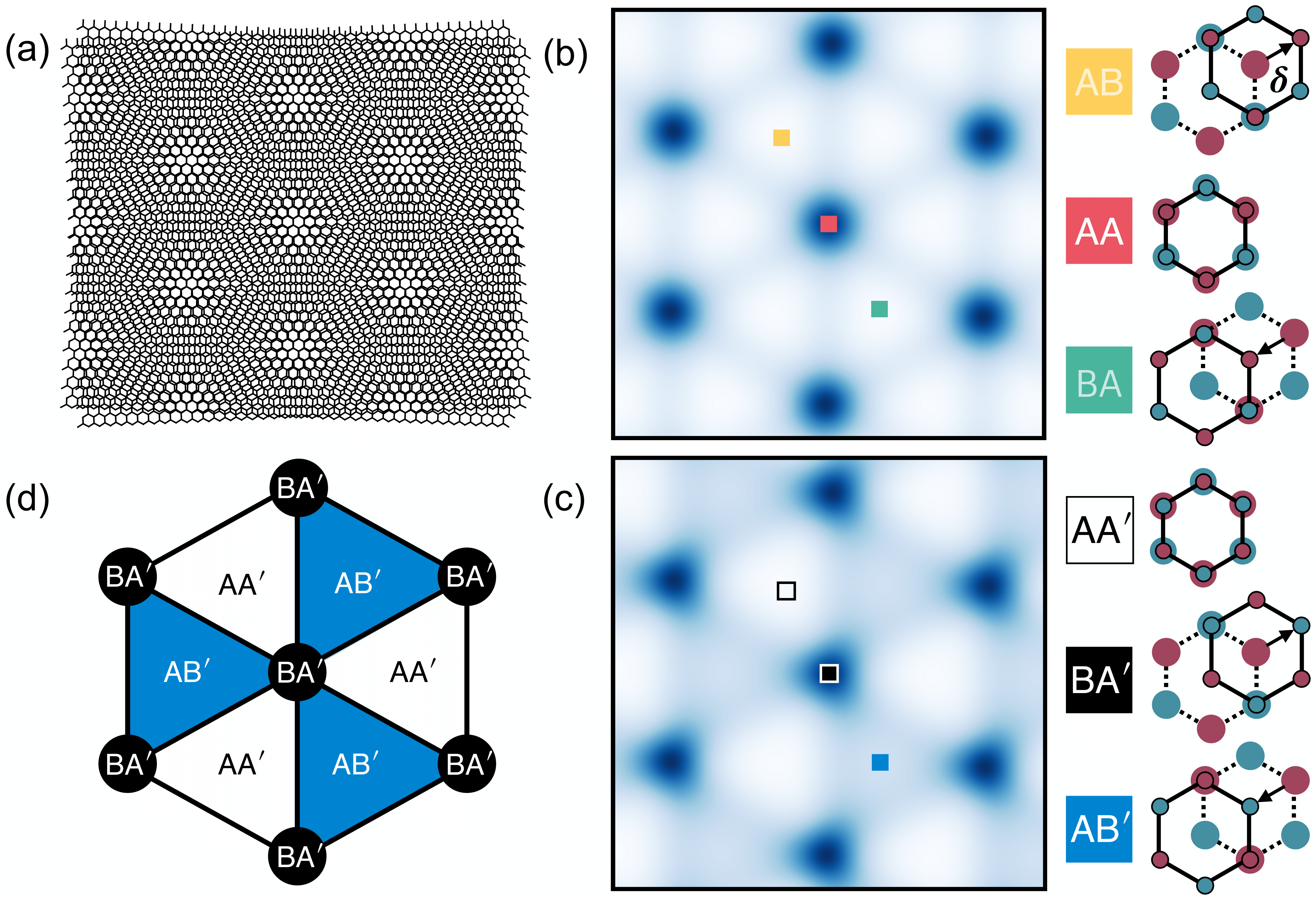}
\caption{(a) Moir\'e pattern of a twisted bilayer h-BN. (b),(c) The interlayer binding potential of the relaxed moir\'e structure formed by twisting the $0^\circ$ and $180^\circ$ configurations, respectively.  The colored squares indicate the respective local stackings within the moir\'e unit cell. The solid lines and smaller circles indicate the top layer, the dashed lines and bigger circles indicate the bottom layer. (d) Schematics of the relaxed moir\'e superlattice for the $180^\circ$ configuration.}
\label{Fig:h-BN}
\end{figure}

Next we consider the interlayer binding energy.  The moir\'e superlattice can be viewed as a periodic modulation of the local atomic registry, characterized by the relative in-plane displacement $\bm\delta(\bm r)$ between the top and bottom layer.  For twisted homobilayers with twist angle $\theta$, $\bm\delta(\bm r) = [R(\theta) - 1]\bm r + \bm u^{(2)}(\bm r) - \bm u^{(1)}(\bm r)$, where $R(\theta)$ is the rotation matrix.  Within the continuum approximation, the interlayer binding energy can be written as~\cite{carr2018,koshino2019,ochoa2019} 
\begin{equation}
U_B = \int d^2r\, V[\bm \delta(\bm r)] \;,
\end{equation}
where $V[\bm\delta]$ is the binding energy density of the rigidly shifted bilayer with shift vector $\bm\delta$.  $V[\bm\delta(\bm r)]$ must satisfy the following symmetry constraints.  First, moir\'e patterns created by honeycomb structures has at least three-fold symmetry (Figure~\ref{Fig:h-BN}).  Secondly, $V[\bm\delta]$ must remain invariant under a constant displacement of one layer (relative to the other) by an atomic lattice vector.  To the lowest order, the Fourier expansion of $V[\bm\delta(\bm r)]$ compatible with three-fold symmetry reads
\begin{equation}
V[\bm\delta(\bm r)] = V_0 \sum_{j=1}^3 
\cos[\bm a_j^* \cdot \bm\delta(\bm r) + \phi] \;,
\end{equation}
where $\bm{a}_j^*$ are the set of first stars of the reciprocal lattice vectors of the honeycomb structure related by $C_{3z}$.  Crucially, $C_{3z}$ allows a finite phase $\phi$, which, except when $\phi = 2n\pi/3$ or $n\pi$ for integer $n$, breaks $C_{2z}$ symmetry and gives rise to chiral phonons.

Armed with the insight that it is the interlayer binding energy that determines the appearance of chiral phonons, we analyze the symmetry of different types of van der Waals heterostructures. 
{We first consider twisted bilayer graphene. The AA (carbon on carbon) stacking has the highest energy.  The local structures at $\bm r $ and $-\bm r$ with respect to the location of the $AA$ stacking are exactly the same and are therefore degenerate in stacking energy. This can be seen from Fig.~\ref{Fig:h-BN}(b) by considering the AB and BA stackings where both sublattices are carbon atoms.  In this case, the interlayer binding energy has $C_{2z}$ symmetry and thus, twisted bilayer graphene cannot host moir\'e chiral phonons.}
We {then} consider a twisted homobilayer made of hexagonal-Boron Nitride (h-BN).  Since a monolayer h-BN breaks inversion symmetry, the top layer can be initially placed at $0^\circ $ (AA stacking) or $180^\circ$ (AA$^\prime$ stacking) with respect to the bottom layer and then twisted to form two different moir\'e superlattices.  For the $0^\circ$ orientation, even though the local structures at $\bm r$ and $-\bm r$ are no longer identical, they are $M_z$ images of each other, therefore the stacking potential still respects $C_{2z}$ symmetry.  This can be seen, for example, by considering again the AB and BA stacking, as marked by yellow and {green} squares in Figure~\ref{Fig:h-BN}(b).  Under $M_z$, the AB and BA stacking transform into each other, therefore they must have the same stacking energy. For the $180^\circ$ configuration the local structures at $\bm r$ and $-\bm r$ are not related by $M_z$ anymore.  Figure~\ref{Fig:h-BN}(c) shows that the AB$^\prime$ and BA$^\prime$ stackings are not $M_z$ images.  {In fact, the AA$^\prime$ stacking is the energy minimum and BA$^\prime$ is the energy maximum~\cite{PhysRevLett.111.036104}}. In this case, the interlayer binding energy break{s} $C_{2z}$ symmetry, {permitting} the appearance of chiral phonons. The same symmetry analysis also applies to twisted transition metal dichalcogenides (TMDs){\cite{PhysRevLett.124.206101}}.
 
Finally, we consider heterobilayers such as graphene/h-BN or any combination of TMDs.  In this case, the AB${^\prime}$ and BA${^\prime}$ stackings are no longer related by $M_z$, and $C_{2z}$ is broken irrespective of the initial orientation. Therefore, chiral phonons must be a possibility.  In addition, the breaking of $M_z$ symmetry can lead to interesting layer-dependent phenomena that will be explored elsewhere.

As a specific example, we calculate the phonon spectrum for a twisted homobilayer.  The Lagrangian is written as
\begin{equation}
\mathcal L = K- U_E - U_B \;,
\end{equation}
where $K = (1/2) \int d^2r \, \sum_\ell \rho^{(\ell)}  (\dot{\bm{u}}^{(\ell)})^2$ is the kinetic energy of the system.  It is convenient to change the basis to symmetric and antisymmetric modes defined as $\bm u^\pm=\bm u^{(1)}\pm \bm u^{(2)}$.  For twisted homobilayers, the moir\'e potential will only affect the $\bm u^-$ mode.  We express $\bm u^-(\bm r,t) = \bm u^-_0(\bm r) + \delta\bm u^-(\bm r,t)$, where $\bm u^-_0$ is the static part describing lattice relaxation and $\delta\bm u^-$ is the dynamical part describing phonons.  The {calculation} follows the procedure outlined in Ref.~\cite{koshino2019} and the details can be found in the Supplementary Material~\cite{supp}.   

\begin{figure}
\includegraphics[width=\columnwidth]{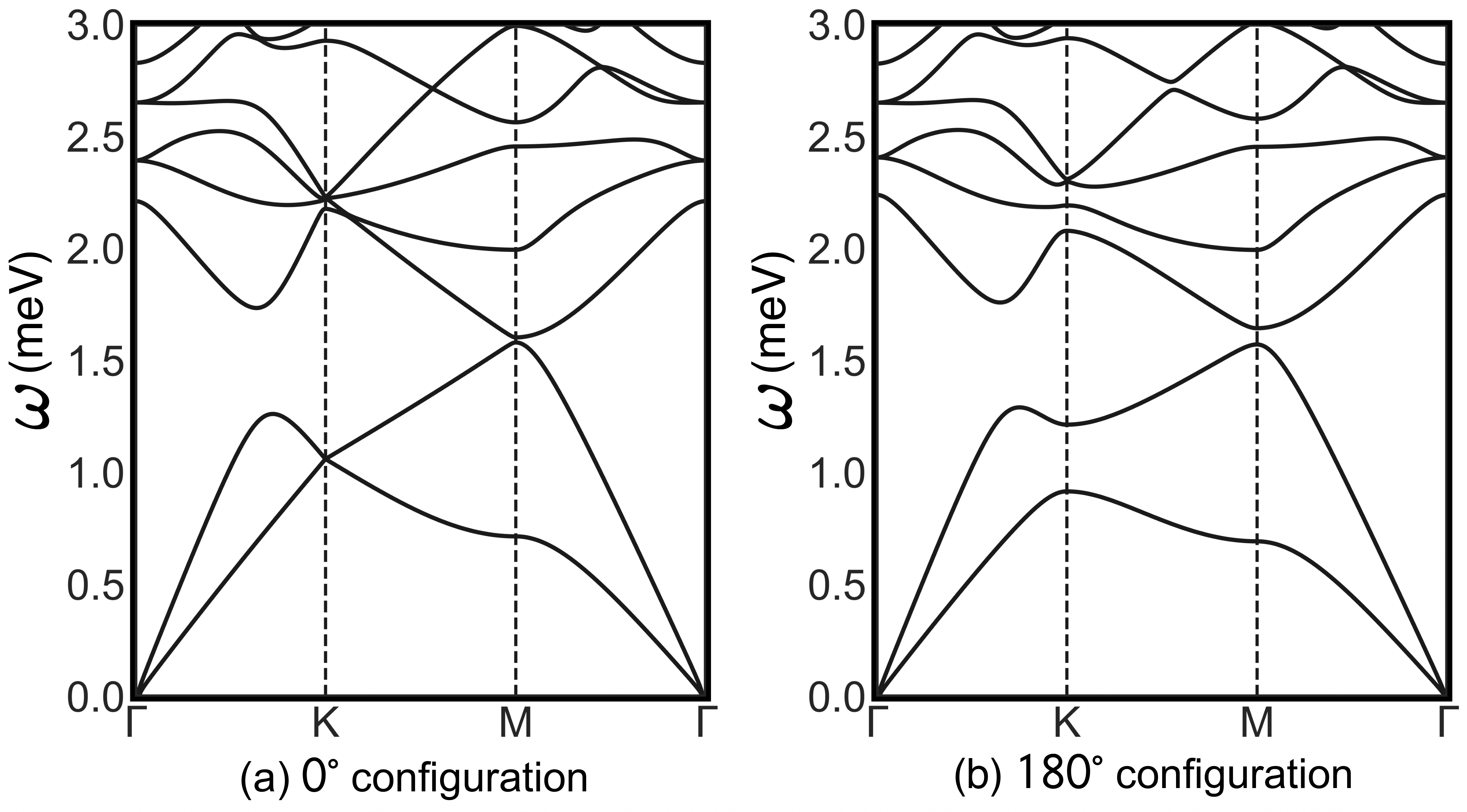}
\caption{The moir\'e phonon band structures at twist angle $\theta = 2^\circ$ for (a) the 0$^\circ$ configuration and (b) the 180$^\circ$ configuration.  The lowest two bands are degenerate at the $\bm K$ point in (a) while in (b) this degeneracy is lifted.}
\label{fig:Fig2}
\end{figure}

We show the calculation of twisted MoS$_2$ for both the $0^\circ$ and $180^\circ$ configurations with a $2^\circ$ twist angle.  For MoS$_2$, the density $\rho = 3.026 \times 10^{-6}$ kg/m$^2$ and Lam\'e coefficients are $\lambda = 3.3$ eV/\AA$^2$ and $\mu=3.6$ eV/\AA$^2$.  The binding energy of the $180^\circ$ configuration are specified by $V_0 = 1.4$~meV and $\phi = -11.04^\circ$~\cite{carr2018}.  For the $0^\circ$ configuration we use the same $V_0$ but set $\phi = 0$.  We first relax the lattice structure, which leads to the formation of large stacking domains.  The resulting interlayer binding potentials are shown in Figure~\ref{Fig:h-BN}(b) and (c) for the 0$^\circ$ and the 180$^\circ$ configuration, respectively.  It can be seen that the 0$^\circ$ configuration preserves $C_{2z}$ while the 180$^\circ$ configuration breaks it.  This  symmetry difference is reflected in the phonon band structures as shown in Figure~\ref{fig:Fig2}.  The two lowest bands are degenerate at the $\bm K$ point for the 0$^\circ$ configuration but split up for the 180$^\circ$ configuration due to the breaking of $C_{2z}$, signifying the appearance of chiral phonons.

\begin{figure}
\includegraphics[width=\columnwidth]{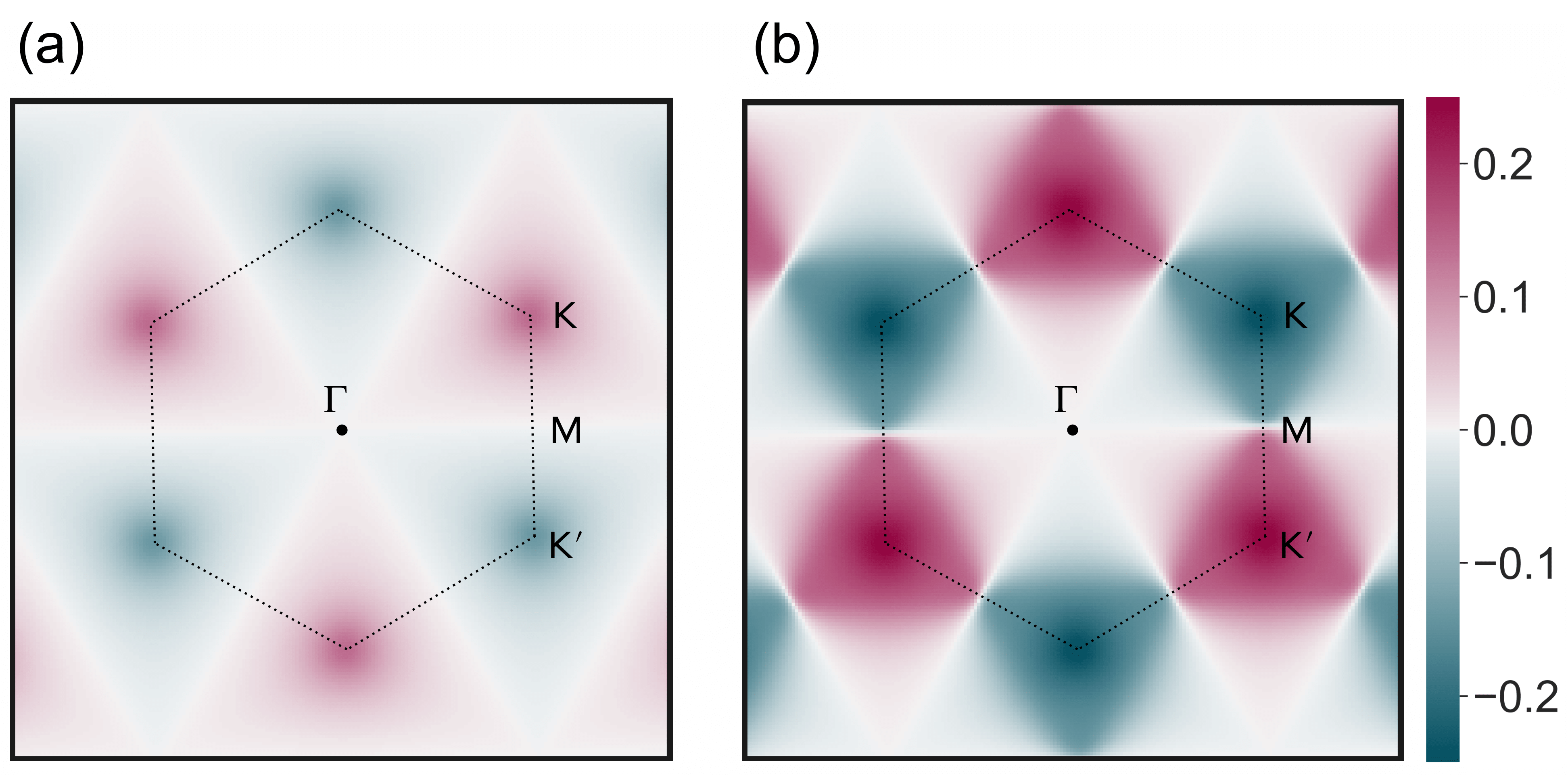}
\caption{The angular momentum {($\hbar$)} of (a) the lowest and (b) the second lowest phonon band in the reciprocal space for the 180$^\circ$ twisted bilayer MoS$_2$ at twist angle $\theta = 2^\circ$.}
\label{fig:Fig3}
\end{figure}

To verify that phonons of the 180$^\circ$ configuration are indeed chiral, we calculate their angular momentum, defined at momentum $\bm q$ as~\cite{supp, zhang2014}
\begin{equation}
L^z_{{\sigma}\bm q} = {\rho} \int d^2r\, \sum_{\ell=1}^2 (\delta \bm u^{(\ell)}_{{\sigma}\bm q} \times \delta \dot{\bm u}^{(\ell)}_{{\sigma}\bm q})_z \;,
\end{equation}
where $u_{\sigma\bm q}$ is the eigenmode of the $\sigma$th branch.
Figure~\ref{fig:Fig3} shows the angular momentum of the two lowest bands in the reciprocal space.  We see that the angular momentum has opposite sign at opposite momentum, as dictated by time-reversal symmetry.  The angular momentum of the lowest band is more localized around the corners of the Brillouin zone compared to the second lowest band, due to the latter's proximity to the third band in energy.  We have also calculated the phonon Berry curvature~\cite{supp}, which shows the same symmetry property as $L^z_{\bm q}$ and can lead to a phonon angular momentum Hall effect.

\begin{figure}
\includegraphics[width=\columnwidth]{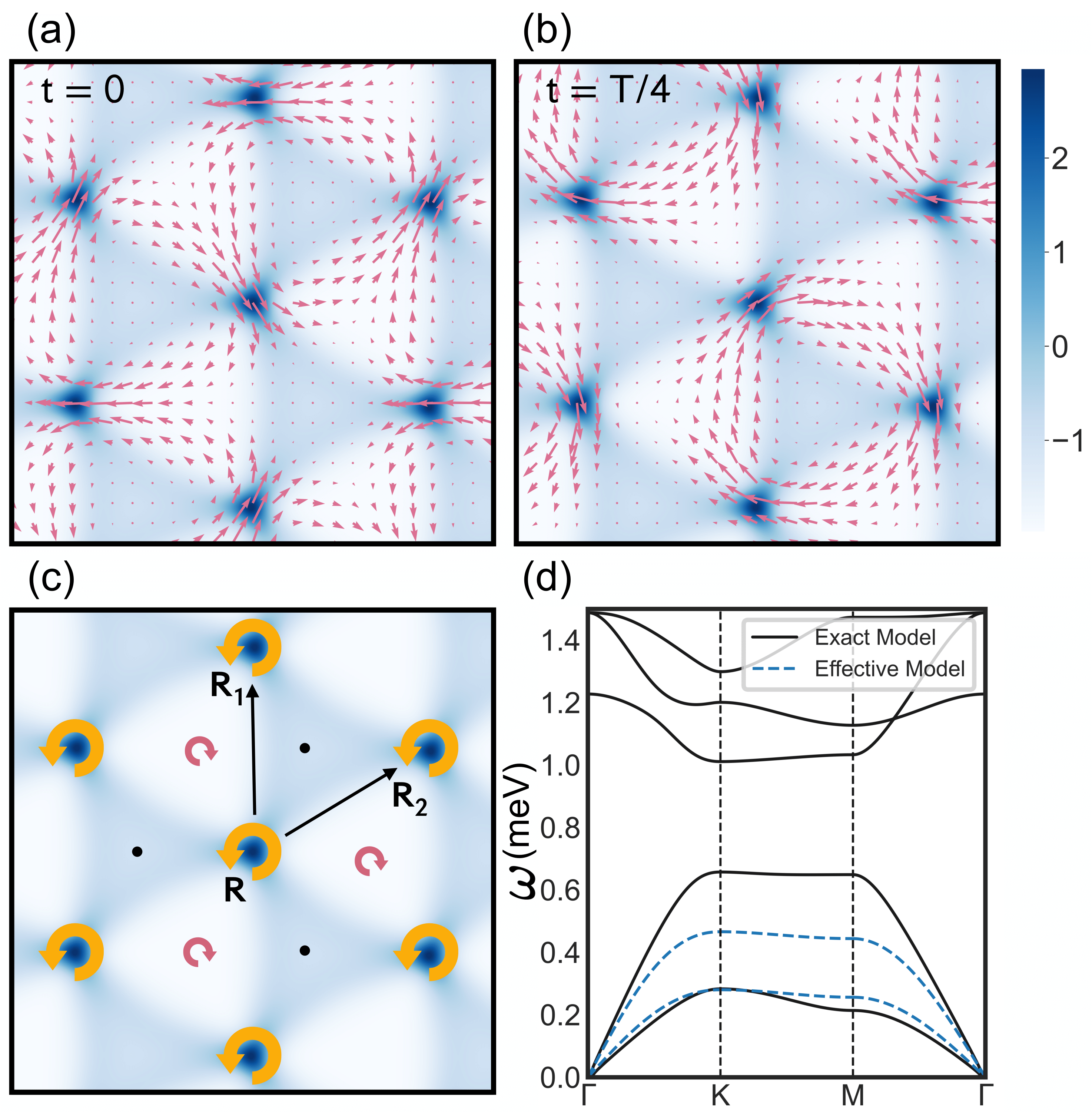}      
\caption{The real space displacement field $\delta\bm u^-(\bm r, t)$ of the lowest band at the $\bm{K}$ point at times (a) $t = 0$ and (b) $t = T$/4 respectively, where $T$ is the time period. The twist angle is 1$^\circ$ for the 180$^\circ$ configuration. The background is the interlayer binding potential of the relaxed structure. The dark blue regions are {BA$^\prime$}, the bigger white regions are {AA$^\prime$} and the smaller light blue regions are {AB$^\prime$}. (c) Schematics of phonon modes show the rotation of different stacking regions. The size of arrows represents the amplitude and the black dots represent the lack of motion.  (d) The band structure for the low energy effective model compared with the exact calculation.}
\label{fig:Fig4}
\end{figure}

To gain further insight into the formation of chiral phonons, we plot $\delta\bm u^-(\bm r, t)$ of the lowest phonon mode at the $\bm K$ point in Figure~\ref{fig:Fig4}(a) and (b).  We see that the {BA$^\prime$} regions rotate anticlockwise while the {AA$^\prime$} regions rotate clockwise with a much smaller amplitude, and the {AB$^\prime$} regions remain stationary. As a result, the phonon mode carries a net angular momentum.  For the next band, we find opposite behavior: the {BA$^\prime$} regions rotate clockwise; the {AA$^\prime$} regions remain stationary and the {AB$^\prime$} regions rotate anticlockwise with a much smaller amplitude. A movie showing their full motion in time can be found in the Supplementary Material~\cite{supp}.

Based on the real-space picture, we develop a low-energy effective model to provide an intuitive explanation for the emergence of chiral phonons.
Since the angular momentum is mainly carried by the motion of the {BA$^\prime$} regions, the chiral phonon mode can be qualitatively understood by taking the {BA$^\prime$} regions as oscillating `points' with displacements $\tilde{\bm u}(\bm{R})$.  The {BA$^\prime$} oscillators are connected by domain walls that separate the {AA$^\prime$} and { AB$^\prime$} domains as shown in Figure~\ref{fig:Fig4}(c).  The potential energy of the system comes from two parts: the domain walls and the domains.  The domain wall energy has been considered previously by Koshino and Son~\cite{koshino2019}; it is written as $E_\text{DW} = (1/2) \sum_{\bm q} \tilde{\bm u}_{-\bm q}^{T} D^\text{DW}(\bm q) \tilde{\bm u}_{\bm q}$, where the dynamical matrix $D^\text{DW}(\bm q)$ was derived in Ref.~\citenum{koshino2019} and reproduced in the Supplementary Material~\cite{supp}.  It satisfies the property $D^\text{DW}(\bm q)=D^\text{DW}(-\bm q)$, i.e., $E_\text{DW}$ does not break $C_{2z}$.  The key to understand the emergence of chirality lies in the potential energy associated with the domains.  During the oscillation of the domain wall network, the areas of stacking domains {AA$^\prime$} and {AB$^\prime$} will change.  This leads to an additional change of the potential energy given by $E_{\text{Domain}} = V_{{AA^\prime}} \Delta A_{{AA^\prime}}  + V_{{AB^\prime}} \Delta A_{{AB^\prime}}$, where $V$ and $\Delta A$ are the average potential and change in area of the respective domains. Let us consider a triangle defined by the three points $\bm R$, $\bm R_1$ and $\bm R_2$ shown in Figure~\ref{fig:Fig4}(c).  Simple geometric consideration yields the change in area as $\Delta A = (1/2) [ (\tilde{\bm{u}}(\bm{R_1})- \tilde{\bm{u}}(\bm{R})) \bm{\times} (\tilde{\bm{u}}(\bm{R_2})- \tilde{\bm{u}}(\bm{R}))] \cdot \bm{z}$, in which we only kept bilinear terms in displacement field as the linear terms average out to zero. Transforming to Fourier space, the potential energy associated with domains reads~\cite{supp}
\begin{equation}
E_\text{Domain} = \alpha \frac{(V_{{AA^\prime}} - V_{{AB^\prime}})}{2} 
\sum_{\bm q} \tilde{\bm u}_{-\bm q}^T 
\begin{bmatrix} 0 & \gamma(\bm q) \\ -\gamma(\bm q) & 0 \end{bmatrix}
\tilde{\bm u}_{\bm q} \;,
\end{equation}
where $\gamma(\bm q) = 4i \sin[\bm q\cdot (\bm L_2 - \bm L_1) / 2] \sin(\bm q \cdot \bm L_1 / 2) \sin(\bm q \cdot \bm L_2 / 2)$, $\alpha$ is an overall fitting parameter to match the energy, and $\bm L_1$ and $\bm L_2$ are moir\'e lattice vectors. This term is proportional to the stacking energy difference between the {AA$^\prime$} and {AB$^\prime$} domains and, therefore, owes its origin to the asymmetric interlayer binding potential.  It breaks $C_{2z}$, which opens the band gap at the $\bm K$ point and gives rise to chiral phonons.  We solve the effective model~\cite{supp} and plot the phonon band dispersion together with that from the elastic theory calculation in Figure~\ref{fig:Fig4}(d).  The agreement is remarkably well given the simplicity of the effective model.  

\begin{figure}
\includegraphics[width=\columnwidth]{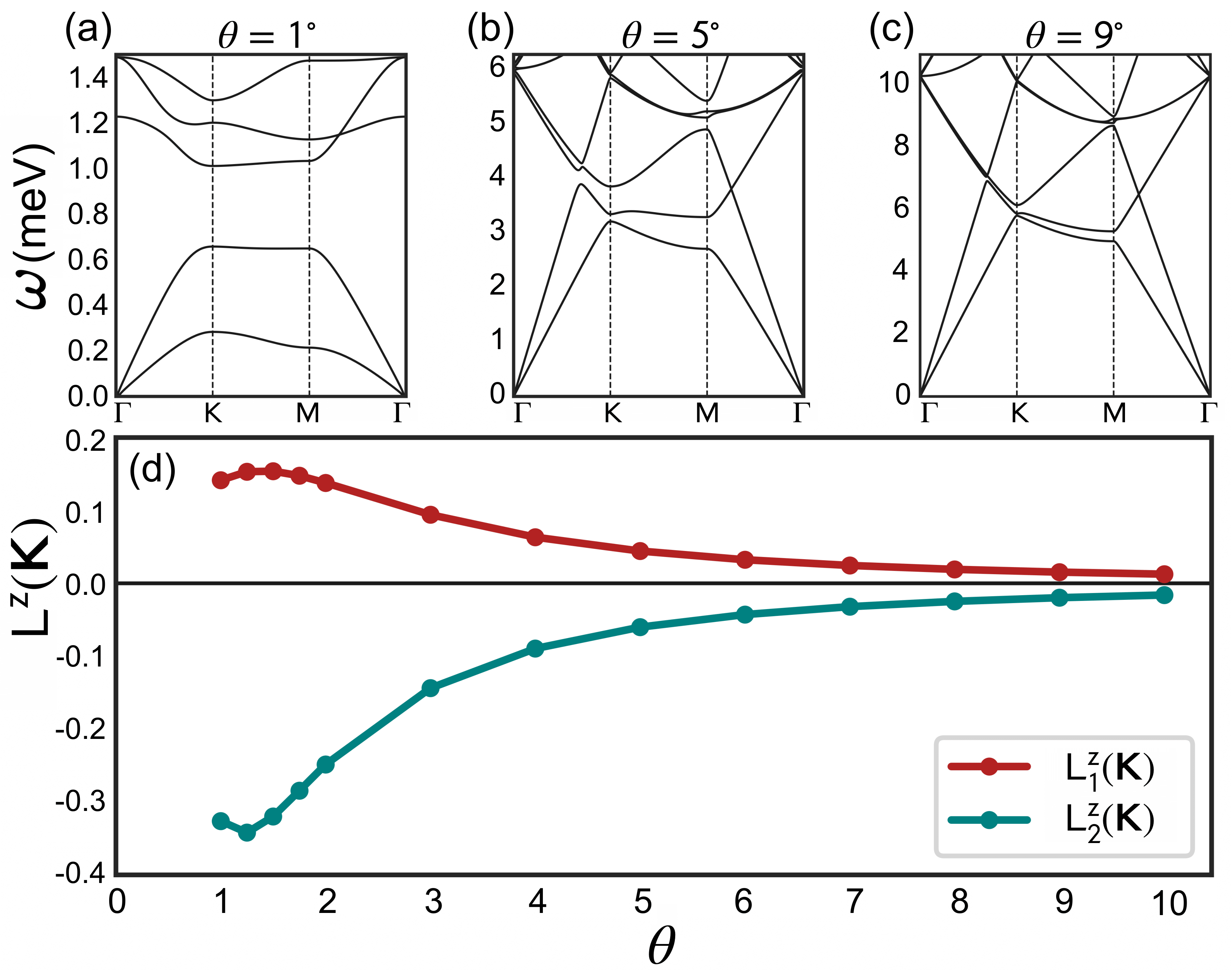}
\caption{(a)-(c) The moir\'e phonon band structures at various twist angles for twisted bilayer MoS$_2$ in the 180$^\circ$ configuration. (d) The angular momentum {($\hbar$)} of the lowest two bands at the $\bm{K}$ point as a function of twist angle.}
\label{fig:Fig5}
\end{figure}

Finally, we show in Figure~\ref{fig:Fig5} the tunability of chiral phonons with respect to the twist angle $\theta$. {As $\theta$ decreases, the BA$^\prime$ regions responsible for larger part of the angular momentum decrease in area. The AA$^\prime$ and AB$^\prime$ regions responsible for smaller counter angular momentum in the first two bands respectively increase in area. Moreover, the area of AA$^\prime$ regions increase more than the AB$^\prime$ regions due to the former's lower stacking energy. This leads to a non-monotonic and asymmetric dependence of angular momentum of the lowest two bands on twist angle.} In addition, the second-lowest band is not entirely separate from the bands above it and therefore its angular momentum does not cancel the angular momentum of the lowest band exactly.

In summary, we showed that chiral phonons can emerge at the moir\'e wavelength in van der Waals heterostructures due to $C_{2z}$ symmetry breaking of the interlayer binding energy.  We used elastic theory to calculate the phonon spectrum for twisted bilayer MoS$_2$ and verified the chiral nature of these phonons by calculating their angular momentum.  The formation of the chiral phonons can be qualitatively understood using an effective model, which emphasizes their origin in the energy difference between the two stacking domains.  Compared to the original proposal of chiral phonons in monolayer TMDs~\cite{zhang2015}, the moir\'e chiral phonons have high tunability, excitation energies in only a few meV, superlattice scale wavelengths and can possibly be mechanically excited. They potentially open new possibilities in phononic twistronic devices.


\begin{acknowledgements}

We are grateful to Hector Ochoa and Haidan Wen for fruitful discussions.  This work was supported by the Department of Energy, Basic Energy Sciences, Materials Sciences and Engineering Division, Pro-QM EFRC (DE-SC0019443).  C.W.\ acknowledges the support from the Department of Energy, Basic Energy Sciences, Materials Sciences and Engineering Division (DE-SC0012509).

\end{acknowledgements}

\textit{Note added.}---Upon the completion of our work, we become aware of a recent preprint in which chiral phonons in moir\'e superlattices was discussed using {  interatomic force-field based simulations}~\cite{maity2021}.

\end{document}


\preprint{APS/123-QED}

\author{Nishchay Suri}
\affiliation{Department of Physics, Carnegie Mellon University, Pittsburgh, Pennsylvania 15213, USA}

\author{Chong Wang}
\affiliation{Department of Materials Science and Engineering, University of Washington, Seattle, Washington 98195, USA}

\author{Yinhan Zhang}
\affiliation{Department of Physics, Carnegie Mellon University, Pittsburgh, Pennsylvania 15213, USA}

\author{Di Xiao}
\affiliation{Department of Materials Science and Engineering, University of Washington, Seattle, Washington 98195, USA}
\affiliation{Department of Physics, University of Washington, Seattle, Washington 98195, USA}
\email{dixiao@uw.edu}
\title{Supporting Information: Chiral Phonons in Moir\'e Superlattices} 

\date{\today}

\maketitle

\section*{Phonons in Moir\'e Superlattices}
In this section we present a detailed derivation of the moir\'e phonon band structures, following Ref.~\citenum{koshino2019} and \citenum{ochoa2019}.  Let us consider the case of twisted homobilayers. We define the primitive lattice vectors of layer 1 as
\begin{equation}
\bm{a}_1=a(1,0) \;, \quad \bm{a}_2 = a(1/2,\sqrt{3}/2) \;.
\end{equation} 
The reciprocal lattice vectors of layer 1 are given as
\begin{equation}
\bm{a}_1^*=2\pi/a(1,-1/\sqrt{3})\;, \quad \bm{a}_2^*=2\pi/a(0,2/\sqrt{3}) \;.
\end{equation}
As a result of the twist, an atom on layer 2 originally located at $\bm{r}_0$ is moved to $\bm{r}={R}(\theta)\bm{r}_0$, where $R(\theta)$ is the rotation matrix.  Then, we define the interlayer atomic shift given by $\bm{\delta}(\bm{r})$ as the in-plane position of an atom on layer 2 measured from its counterpart in layer 1. In the case of pure rotation we have
\begin{equation}
\bm{\delta}_0(\bm{r})=\bm{r}-\bm{r}_0=(1-{R}^{-1}(\theta))\bm{r} \;.
\end{equation} 
The moir\'e lattice and reciprocal lattice vectors are given as
\begin{equation}
\bm{L}_i =(1-{R}^{-1}(\theta))^{-1}\bm{a}_i \;, \quad \bm{G}_i = (1-{R}(\theta))\bm{a}_i^* \;,
\end{equation}
respectively.

The total instantaneous interlayer atomic shift is given by $\bm{\delta}(\bm{r})=\bm{\delta}_0(\bm{r})+ \bm{u}^-(\bm{r},t)$ , where $\bm{u}^-=\bm{u}^{(2)}-\bm{u}^{(1)}$ is the relative deformation field. The Lagrangian density is composed of the kinetic energy $K$, intra-layer elastic energy $U_E$ and the inter-layer binding energy $U_B$,
\begin{align}
    \mathcal{L} &= K- U_{E} -U_B\;,\\
    K&=  \int d^2 r \ \frac{\rho}{2} \sum_{\ell=1}^2 (\dot{\bm{u}}^{(\ell)})^2\;, \\
    U_{E}&=\int d^2 r \ \sum_{\ell=1}^2 \bigg\{ \frac{\lambda}{2}\left(\partial_{\alpha} u_{\alpha}^{(\ell)}\right)^{2}+\frac{\mu}{4}\left(\partial_{\alpha} u_{\beta}^{(\ell)}+\partial_{\beta} u_{\alpha}^{(\ell)}\right)^{2} \bigg\}\;,\\
    U_B&=\int d^2 r \ V_0  \sum_{j=1}^3 \cos(\bm{a}_j^* \cdot \bm{\delta}(\bm{r}) + \phi)\;,
\end{align}
where $\ell$ is the layer index, $\rho$ is the density, and $\mu,\lambda$ are the Lam\'e coefficients. Similar to $\bm{u}^-$, it is convenient to define $\bm{u}^+=\bm{u}^{(2)}+\bm{u}^{(1)}$ and write the above energies and equations in these coordinates:
\begin{align}
    K&= \int d^2 r \ \frac{\rho}{4} \sum_{l=1}^2 (\dot{\bm{u}}^{+2} + \dot{\bm{u}}^{-2})\;, \\
    U_{E}&=\int d^2 r \ \frac{1}{2} \bigg\{ \frac{\lambda}{2}\lbrack\left(\partial_{\alpha} u_{\alpha}^{+}\right)^{2} +(\partial_{\alpha} u_{\alpha}^{-})^2 \rbrack +\frac{\mu}{4}\lbrack \left(\partial_{\alpha} u_{\beta}^{+}+\partial_{\beta} u_{\alpha}^{+}\right)^{2}+ \left(\partial_{\alpha} u_{\beta}^{-}+\partial_{\beta} u_{\alpha}^{-}\right)^{2}  \rbrack \bigg\}\;,\\
    U_B&=\int d^2 r \ V_0 \sum_{j=1}^3 \cos(\bm{G}_j \cdot \bm{r} + \bm{a}_j^* \cdot \bm{u}^- + \phi)\;.
\end{align}

Since mirror $z$ symmetry is present in elastic theory, $\bm{u}^+$ and $\bm{u}^-$ modes are completely decoupled. The $\bm{u}^+$ mode is unaffected by the binding energy; it can be ignored as its phonon dispersion will be the zone folded monolayer dispersion. In the following we will only focus on the $\bm u^-$ mode.

We express $\bm{u}^-(\bm{r},t)=\bm{u}^-_0(\bm{r}) + \delta\bm{u}^-(\bm{r},t)$, where $\bm{u}_0$ is the equilibrium part describing the lattice relaxation and $\delta\bm{u}^-$ is the dynamical part describing phonons. Using the Euler-Lagrange equations for $\bm{u}^-_0$, we obtain:
\begin{align}
    \mu \partial_{\beta}^{2} u_{0\alpha}^{-}+(\mu+\lambda) \partial_{\alpha} \partial_{\beta} u_{0\beta}^{-}+2 V_{0} \sum_{j}\left(a_{j}^{*}\right)_{\alpha} \sin \left(\boldsymbol{G}_{j} \cdot \boldsymbol{r}+\boldsymbol{a}_{j}^{*} \cdot \boldsymbol{u}_0^{-} + \phi \right)=0\;.
\end{align}
To solve for the relaxed structure we transform to Fourier space and proceed to solve a system of self-consistent equations. We write $\bm u^-_0 = \sum_{\bm G} \bm u^-_{0\bm G} e^{i\bm G \cdot \bm r}$ where $\bm{G}=m  \bm{G}_1 + n  \bm{G_2}$, such that $m,n \in \mathbb{Z}$.  We note that with enough harmonics we can get the relaxation to converge, leading to an all real phonon spectrum which is physical. For the calculation shown in main text, it suffices to take $m,n \in \{ \pm 1, \dots, \pm 20\}$ for the most strongly relaxed $1^\circ$ twist configuration. Since $\phi$ breaks the six-fold symmetry, we impose $C_3$ constraint on $\bm{u}_0^-$:
\begin{align}
    {R}_{\frac{2\pi}{3}} \bm{u}^-_{0,\bm q}=\bm{u}^-_{0,{R}_{\frac{2\pi}{3}} \bm{q}}\;,
\end{align}
where ${R}_{\frac{2\pi}{3}}$ is the rotation matrix.
It can further be shown that the equation for phonon dynamics is
\begin{align}
    \rho \delta \ddot{u}_{\alpha}^{-}=\mu \partial_{\beta}^{2} \delta u_{\alpha}^{-}+(\mu+\lambda) \partial_{\alpha} \partial_{\beta} \delta u_{\beta}^{-}+2 V_{0} \sum_{j}\left(a_{j}^{*}\right)_{\alpha}\left(a_{j}^{*}\right)_{\beta} \cos \left(\boldsymbol{G}_{j} \cdot \boldsymbol{r}+\boldsymbol{a}_{j}^{*} \cdot \boldsymbol{u}_{0}^{-} + \phi \right) \delta u_{\beta}^{-}\;.
\end{align}
We expand the above equation in Fourier series,
\begin{gather}
    \delta\bm{u}^-(\bm{r},t) = e^{-i\omega t} \sum_{\bm{q}}  \bm{\epsilon}^-(\bm{q})e^{i\bm{q}\cdot \bm{r}}\;,\\
    \cos(\bm{G}_j\cdot \bm{r} + \bm{a}_j^* \cdot \bm{u}_0^- + \phi) = \sum_{\bm{G}} h_{\bm{G}}^j e^{i\bm{G}\cdot \bm{r}}\;, \\
    \rho \omega^2  \ \epsilon^-_\alpha (\bm{q}) = \mu q_\beta^2 \ \epsilon_\alpha^-(\bm{q}) + (\mu+\lambda)q_\alpha q_\beta \ \epsilon_\beta^-(\bm{q})-2 V_0 \sum_{j \bm{G}} (a_j^*)_\alpha (a_j^*)_\beta h_{\bm{G}}^j \ \epsilon^-_\beta(\bm{q}-\bm{G})\;,
\end{gather}
where $\bm{G}=m  \bm{G}_1 + n  \bm{G_2}$ and $m,n \in \mathbb{Z}$. The solution of the above central equation is given by the set of eigenfunctions
\begin{align}
\delta \bm{u}^-_{\sigma \bm q}(\bm r) = e^{-i\omega t}\sum_{\bm G} \bm{\epsilon}^-_{\sigma,\bm q - \bm G} e^{i(\bm q-\bm G)\cdot \bm r}\;,
\label{Sol:Central}
\end{align}
for eigenvalues $\omega_{\sigma}^2(\bm q)$ where $\sigma$ is the band index.

\begin{figure}[h!]
\includegraphics[width=\textwidth]{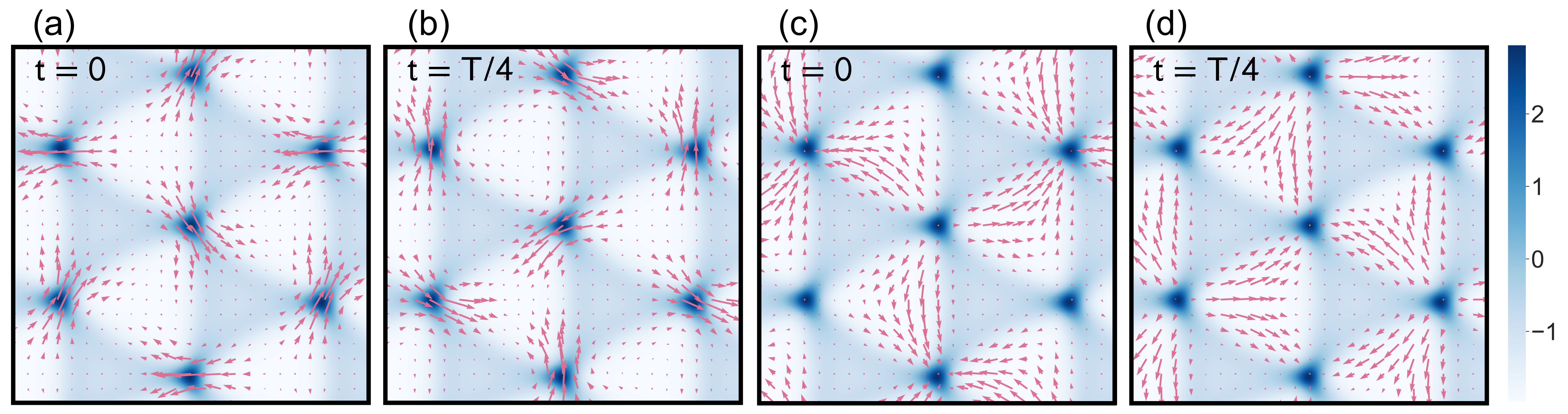}
\caption{The twist angle is 1$^\circ$ for the 180$^\circ$ configuration MoS$_2$. The background is the interlayer binding potential of the relaxed structure. The dark blue regions are BA$^\prime$, the bigger white regions are AA$^\prime$ and the smaller light blue regions are AB$^\prime$. The real space displacement field $\delta\bm u^-(\bm r, t)$ of the second lowest band at the $\bm{K}$ point at times (a) $t = 0$ and (b) $t = T$/4 respectively, where $T$ is the time period. The real space displacement field for the third lowest band at times (c) $t=0$ and (d) $t=T/4$ respectively.}
\label{Fig:RealSpace}
\end{figure}

For the actual calculations we considered twisted bilayer MoS$_2$.  For MoS$_2$, $a=3.18$\AA. 
The density $\rho = 3.026 \times 10^{-6}$ kg/m$^2$. The Lam\'e constants are given by $\lambda=3.3 \ \text{eV}/\text{\AA}^2$, $\mu=3.6 \ \text{eV}/\text{\AA}^2$. The parameters for binding energy are given by Ref.~\citenum{carr2018}:
\begin{align}
\Delta &= \sum_{i=1}^3 \lbrack \Delta_1 \cos(\bm{G}_j\cdot \bm{r}+\bm{a}^*_j\cdot \bm{u}^-) + \Delta_2  \sin(\bm{G}_j\cdot \bm{r}+\bm{a}^*_j\cdot \bm{u}^-) \rbrack/Area \\
&= V_0\sum_{i=1}^3 \cos(\bm{G}_j\cdot \bm{r}+\bm{a}^*_j\cdot \bm{u}^- + \phi)\\
&= V_0\sum_{i=1}^3 (\cos(\bm{G}_j\cdot \bm{r}+\bm{a}^*_j\cdot \bm{u}^-)\cos(\phi) - \sin(\bm{G}_j\cdot \bm{r}+\bm{a}^*_j\cdot \bm{u}^-) \sin(\phi))\;.
\end{align}
Using the relations $V_0 = \sqrt{\Delta_1^2 + \Delta^2_2}/\text{unit cell area}$ and $  \tan\phi= - \Delta_2/\Delta_1$. For $0^\circ$ MoS$_2$, with $\Delta_1=0.014$ eV, $ \Delta_2=0$, we obtain $ V_0=1.6 \ \text{meV/\AA}^2 $ and $  \phi=0^\circ$. For $180^\circ$ MoS$_2$, with $\Delta_1=0.0123$ eV, $ \Delta_2=0.0024$ eV, we obtain $ V_0=1.4 \ \text{meV/\AA}^2$ and $  \phi=-11.04^\circ$.  In our calculation of the phonon spectrum (Fig.~2 in main text), we used the same $V_0 = 1.4$ meV/\AA$^2$ for both configurations to illustrate the effect of inversion symmetry breaking (finite $\phi$). The real space phonon field for the second and third lowest bands are shown in Fig.~\ref{Fig:RealSpace}. To demonstrate the chirality of respective domains, the fields for each band are plotted for times $t=0$ and $t=T/4$, where $T$ is the time period.

\section*{Angular Momentum}
In this section we present the detailed steps on the calculation of phonon angular momentum shown in Fig.~3 and Fig.~5 in the main text.  The phonon angular momentum is defined as~\cite{zhang2015,zhang2014}: 
\begin{align}
L^z = \rho \int d^2 \bm r \sum_{\ell=1,2}( \delta u_x^{(\ell)} \delta\dot{u}_y^{(\ell)} - \delta u_y^{(\ell)} \delta \dot{u}_x^{(\ell)})  = \rho \int d^2 \bm r \ \delta u^T(\bm r) \ i M \ \delta \dot{u}(\bm r)\;,
\end{align}
where $M = I_2 \otimes \sigma_y$ and $\delta u{(\bm{r})}=[\delta u_{x}^{(1)} (\bm r), \delta u_{y}^{(1)}(\bm r), \delta u_{x}^{(2)}(\bm r),\delta u_{y}^{(2)}(\bm r)]^T$. For the unit cell number $n$ and the sublattice atom number $\alpha$, we have
\begin{align}
\delta u_{n,\alpha} = \sum_k \tilde{\epsilon}_{k,\alpha} e^{i ( \bm k  \cdot \bm R_n - \omega_k t)} \sqrt{\frac{\hbar}{2\omega_k m_\alpha N}} a_k + h.c.\;,
\end{align}
where $k=(\bm k, \sigma)$, $\sigma$ is the band number, $m_\alpha$ is the mass of the $\alpha^{\text{th}}$ atom and $N$ is the number of unit cells. The eigenfunction $\tilde{\epsilon}_{k\alpha} = \sum_G \epsilon_{ k \bm G}\ e^{i(\bm k-\bm G)\cdot \bm r_\alpha}$, where $\epsilon_{k \bm G}$ are the coefficients of the block function solution to the central equation. The $\epsilon_{k,\bm G}$ can be obtained from Eq.~\ref{Sol:Central} by setting $\bm u^+=0$ as we are calculating angular momentum for the $\bm u^-$ bands. The $\epsilon_{k\bm G}$ are normalized with respect to harmonics $\bm G$. The vector $\bm r_\alpha$ points to the $\alpha^{\text{th}}$ atom in moir\'e unit cell. In the case where all sublattices are the same, $m_\alpha = \rho \frac{A_u}{N_u}$, where $A_u$ is the area of the unit cell and $N_u$ are the number of atoms in the unit cell. Using $\bm{G}\cdot \bm{R}_n = 2n\pi$, we obtain
\begin{align}
\delta u(\bm r) = \frac{1}{\sqrt{N_u}}\sum_{k\bm G} \epsilon_{k \bm G} e^{i ((\bm k-\bm G) \cdot \bm r - \omega_k t)} \sqrt{\frac{\hbar N_u}{2\omega_k \rho A_u N}} a_k + h.c.\;,
\end{align}
where $\bm r$ is over the entire lattice. Writing the angular momentum
\begin{align}
L^z = \frac{\hbar}{2A_uN}\int d^2 \bm r \sum_{k k^\prime\bm G \bm G^\prime} &\bigg( \epsilon_{k \bm G}^\dagger M \epsilon_{k^\prime \bm G^\prime} \sqrt{\frac{\omega_{k^\prime}}{\omega_{k}}} a_k^\dagger a_{k^\prime}\nonumber \\&+ \epsilon_{k^\prime \bm G^\prime}^T (-M) \epsilon^*_{k \bm G} \sqrt{\frac{\omega_{k}}{\omega_{k^\prime}}} a_{k^\prime} a^\dagger_{k} \bigg) e^{i[(\bm k^\prime-\bm G^\prime) - (\bm k - \bm G)]\cdot \bm r} e^{i(\omega_k-\omega_{k^\prime})t}\;,
\end{align}
where we ignore the fast moving $aa$ and $a^\dagger a^\dagger$ terms. Since $\bm k$ runs only in the first brillouin zone and $\bm G$ are the reciprocal lattice vectors we have $\int d^2 \bm r \ e^{i[\bm k^\prime-\bm G^\prime - (\bm k - \bm G)]\cdot \bm r} = (2\pi)^2 \delta (\bm k^\prime-\bm k -\bm G^\prime + \bm G) \rightarrow 
{\delta_{\bm k,\bm k^\prime}\delta_{\bm G \bm G^\prime}N A_u} $. Therefore
\begin{align}
L^z = \frac{\hbar}{2}  \sum_{k \bm G } \bigg( \epsilon_{k \bm G}^\dagger M \epsilon_{k \bm G} \ a_k^\dagger a_{k} + \epsilon_{k \bm G}^T (-M) \epsilon^*_{k \bm G} \  a_{k} a^\dagger_{k}    \bigg)\;.
\end{align}
 We use the properties $\bm{\epsilon}_{k\bm G}^T (-M) \bm{\epsilon}_{k\bm G}^* = \bm{\epsilon}_{k\bm G}^\dagger M \bm \epsilon_{k \bm G}$ and $[a_k,a^\dagger_k]=1$ to obtain 
\begin{align}
{L^z}=   \sum_{k} {L^z_k} \bigg(a_k^\dagger a_k + \frac{1}{2}\bigg)\;.
\end{align}
In the main text we plot angular momentum at $k=(\bm{k},\sigma)$ given by $L^z_k = \sum_{\bm G} (\epsilon_{k\bm G}^\dagger M \epsilon_{k \bm G} ) \hbar$.

\section*{Berry Curvature}

\begin{figure}[h!]
\centering
\includegraphics[scale=0.25]{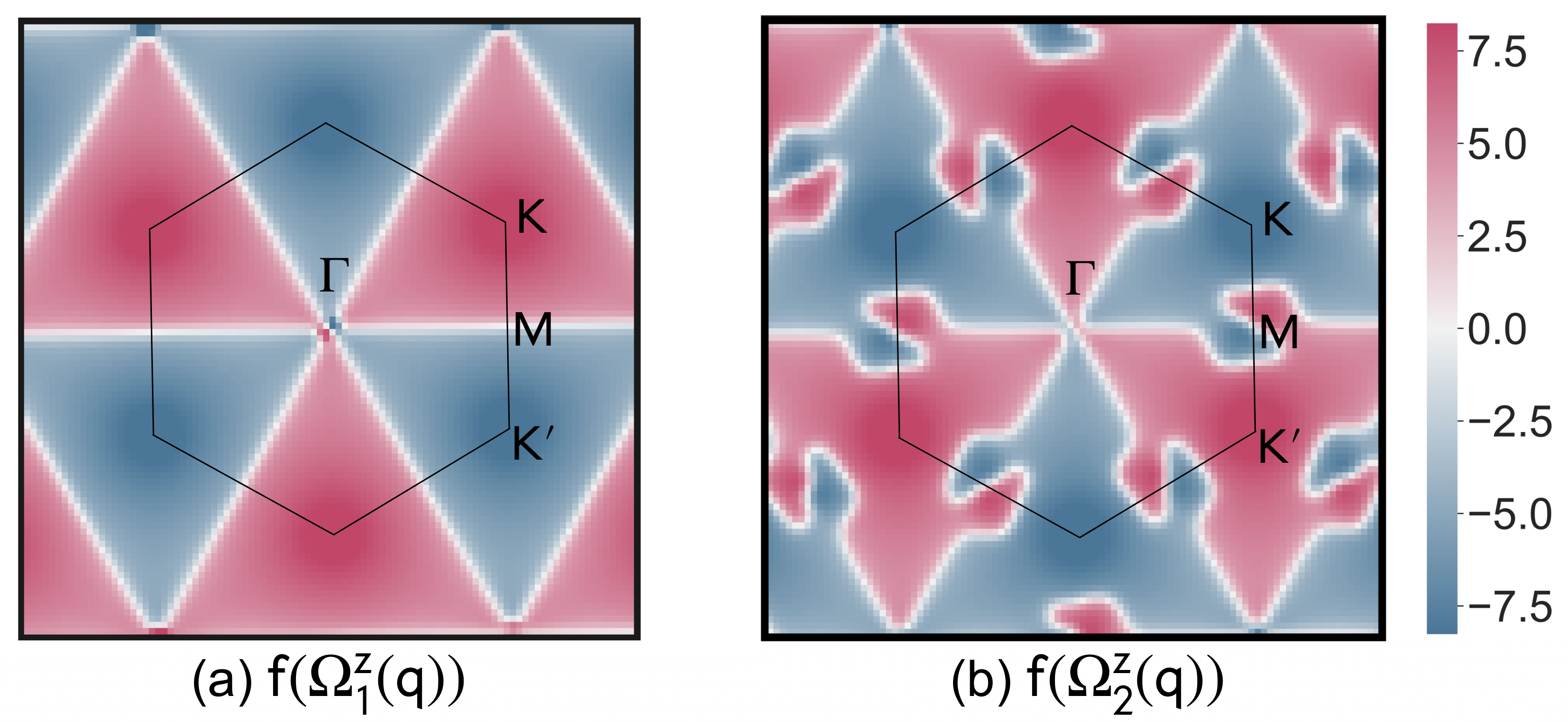}
\caption{The log scaled Berry curvature of (a) the lowest and (b) the second lowest phonon band in the reciprocal space for the 180$^\circ$ twisted bilayer MoS$_2$ at twist angle $\theta = 2^\circ$. The scaling function is $f(\Omega^z) = \text{sign}(\Omega^z) \ln(1+|\Omega^z|)$ }
\label{fig:Berry}
\end{figure}

We follow the procedure in Ref.\citenum{zhang2019thermal} to calculate the phonon Berry curvature. We obtain the Hamiltonian from the Lagrangian using Legendre transformation and write it in Fourier space $  H = \frac{1}{2} \sum_q \psi^\dagger_{\bm{q}} H_{\bm{q}} \psi_{\bm{q}} $. We define the basis as $\psi_{\bm{q}}=\lbrack \delta \tilde{u}^{-}_{({\bm{q}-\bm{G}})x}, \delta \tilde{u}^-_{({\bm{q}-\bm{G}})y}, \tilde{p}^-_{-({\bm{q}-\bm{G}})x}, \tilde{p}^-_{-({\bm{q}-\bm{G}})y}\rbrack^T$ where each of the four operators $\delta \tilde{u}^-_{({\bm{q}-\bm{G}})\alpha}$, $ \tilde{p}^-_{({\bm{q}-\bm{G}})\alpha}$ are $N$ dimensional, with $\bm{G}$ taking $N$ values given by $\bm{G}=m_1 \bm{G}_1 + m_2 \bm{G}_2$ s.t. $m_1,m_2 \in \mathbb{Z}$. Here, we choose an appropriately large cutoff $N$, which is sufficient for convergence. The $4N \times 4N$ matrix $H_{\bm{q}}$ is given as 
\begin{align}
H_{\bm{q}} = \frac{1}{2}\begin{bmatrix}
        K_{\bm{q}} + V_{\bm{q}} & 0 \\ 0 & \frac{\rho}{4} I_{2N}
        \end{bmatrix}\;,
\end{align}
where $K_q$ is the spring constant matrix, given as 
\begin{align}K_{\bm{q}}=
    \begin{bmatrix}
       \text{diag}\{(\lambda+2\mu) q^2_x + \mu q_y^2 \}& \text{diag}\{(\lambda + \mu) q_x q_y\}\\
        \text{diag}\{(\lambda + \mu) q_x q_y\} & \text{diag}\{(\lambda+2\mu) q^2_y + \mu q_x^2\}
    \end{bmatrix}\;,
\end{align} 
where each element is a $N$ dimensional diagonal matrix $\text{diag}\{{q}\} = \text{diag} (q-G_0, \dots ,q-G_{N-1})$. Similarly the interlayer interaction enters as
\begin{align}
V_{\bm{q}}= -2 V_0 \sum_j 
    \begin{bmatrix} 
    a_{jx}^{*2} \ h^j & a_{jx}^* a_{jy}^* \ h^j   \\ a_{jx}^* a_{jy}^* \ h^j & a_{jy}^{*2} \ h^j
    \end{bmatrix}\;,                                                    
\end{align} 
where the elements of matrix $h^j$ are given as $(h^j)_{\bm{G},\bm{G}^\prime} = h^j_{\bm{G}-\bm{G}^\prime}$. The element $h^j_{\bm{G}}$ is calculated as $\cos(\bm{G}_j\cdot \bm{r} + \bm{a}_j^* \cdot \bm{u}_0^- + \phi) = \sum_{\bm{G}} h_{\bm{G}}^j e^{i\bm{G}\cdot \bm{r}}$.
Using the Heisenberg equation of motion $i \partial_t \psi_{\bm{q}} = \lbrack \psi_{\bm{q}} , H \rbrack$, we obtain $i \mathcal{J} \partial_t \psi_{\bm{q}} = H_{\bm{q}} \psi_{\bm{q}}$. The matrix $\mathcal{J}=\lbrack \psi_{\bm{q}} , \psi_{\bm{q}}^\dagger\rbrack $ is given as
\begin{align}
    \mathcal{J} = 2i \begin{bmatrix} 0 & I_{2N} \\ -I_{2N} & 0 \end{bmatrix}\;.
\end{align} 

We solve the eigenvalue problem $\omega_{n\bm{q}} \mathcal{J} \psi_{n\bm{q}} = H_{\bm{q}} \psi_{n{\bm{q}}}$. The Berry curvature defined as $\Omega_{n}^{z}=\partial_{q_{x}} A_{n y}-\partial_{q_{y}} A_{n x}$, where $\boldsymbol{A}_{n}=i\left\langle\psi_{n \boldsymbol{q}}\left|\mathcal{J} \partial_{\boldsymbol{q}}\right| \psi_{n \boldsymbol{q}}\right\rangle /\langle\psi_{n \boldsymbol{q}}|\mathcal{J}| {\psi}_{n \boldsymbol{q}}\rangle$. 

We show the log-scaled berry curvature in Fig.~\ref{fig:Berry} for the lowest two bands respectively in the reciprocal space. The log function is defined as $f(\Omega^z) = \text{sign}(\Omega^z) \ln(1+|\Omega^z|)$. The symmetry properties are similar to the angular momentum discussed in the main text. The higher band is slightly dissimilar to the lower band: it has another Berry curvature hot spot as a result of anti-crossing point with a higher band at the $\bm{M}$ point [see Fig.~2(b) in the main text].

\section*{Effective Model}

The effective model of our chiral phonons is an extension of the effective model developed for twisted bilayer graphene by Koshino and Son~\cite{koshino2019}.  As we discussed in the main text, there are two contributions to the total potential energy, those associated with domain walls and those associated with domains.  The domain wall potential has been discussed by Koshino and Son.  It is given by~\cite{koshino2019},
\begin{equation}
E_{\text{DW}} = \frac{1}{2} \sum_{\mathbf{q}} \tilde{\mathbf{u}}_{-\mathbf{q}}^{T} {D}^{DW}(\mathbf{q}) \tilde{\mathbf{u}}_{\mathbf{q}}\;,
\end{equation}
where
\begin{equation}
D^{DW}_{\mu \nu}(\mathbf{q}) =\sum_{i=1}^{3} \frac{\alpha V_{0} w_{d}}{2L_{M}}\left(2 \sin \frac{\mathbf{q} \cdot \mathbf{L}_{i}^{M}}{2}\right)^{2}\left[\delta_{\mu \nu}-\frac{\left(\mathbf{L}_{i}^{M}\right)_{\mu}\left(\mathbf{L}_{i}^{M}\right)_{v}}{L_{M}^{2}}\right]\nonumber\;.
\end{equation}
where $w_{d} \approx \frac{a}{4} \sqrt{\frac{2(\lambda+\mu)}{V_{0}}}$ is the width of domain walls, $\bm{L}_i$ are the first star of moir\'e lattice vectors such that $L_M=|\bm{L}_i|$
~\footnote{Koshino and Son in Ref.~\citenum{koshino2019} take $2V_0$ as the binding potential energy constant compared to $V_0$  used in this letter. The expression for $D^{DW}$ has a factor of half to account for the change.}
. 
It satisfies the property ${D}^\text{DW}(\bm q)={D}^\text{DW}(-\bm q)$, i.e., $E_\text{DW}$ does not break $C_{2z}$.

In the main text, we argued that when the interlayer binding energy breaks $C_{2z}$ symmetry, there is an additional potential energy associated with the area change of the stacking domains,
\begin{equation}
    E_{\text{Domain}} = \frac{1}{2} \sum_{\mathbf{q}} \tilde{\mathbf{u}}_{-\mathbf{q}}^{T} D^\text{Domain}(\mathbf{q}) \tilde{\mathbf{u}}_{\mathbf{q}} = \alpha \frac{(V_{AA^\prime}-V_{AB^\prime})}{2} \sum_{\mathbf{q}} \tilde{\mathbf{u}}_{-\mathbf{q}}^{T} \begin{bmatrix} 
    0 & \gamma_{AA^\prime}(\bm{q})\\
    -\gamma_{AA^\prime}(\bm{q}) & 0
    \end{bmatrix}
    \tilde{\mathbf{u}}_{\mathbf{q}}\;,
\end{equation}
where
\begin{equation}
\gamma_{AA^\prime}(\bm{q}) = 4i \sin(\frac{\bm{q}\cdot (\bm{L_2 - L_1})}{2}) \sin(\frac{\bm{q}\cdot\bm{L_1}}{2}) \sin(\frac{\bm{q}\cdot\bm{L_2}}{2})
\end{equation}
with $\bm L_1$ and $\bm L_2$ being the moir\'e lattice vectors.  It must be noted that only bilinear terms in displacement contribute to the change in area as the linear terms average out to zero. For twisted bilayer graphene, the AA regions have the highest energy, therefore $E_{\text{Domain}} \propto V_{AB}-V_{BA}$. Since $V_{AB} = V_{BA}$, this term does not appear in Koshino and Son's model.

Combining $E_\text{DW}$ and $E_\text{Domain}$ together, the equation of motion is given by
\begin{equation}
M \ddot{\bm{\tilde u}}_{\bm q} = -[D^\text{DW} + D^\text{Domain}]\bm{\tilde u}_{\bm q} \;,
\end{equation}
where $M = \rho(a/w_d)^2w_dL_M$ is the effective mass~\cite{koshino2019}.  Solving the above equation yields the phonon band dispersion of the effective model shown in Fig.~4(d) in the main text.


\bibliography{ChiralPhoMoirebib}